# Modeling CoVid-19 Diffusion with Intelligent Computational Techniques Is Not Working. What Are We Doing Wrong?

Marco Roccetti[1], Giovanni Delnevo[1]

[1]Department of Computer Science and Engineering,
Alma Mater Studiorum - University of Bologna
Mura Anteo Zamboni 7, 40127, Bologna Italy
marco.roccetti@unibo.it, giovanni.delnevo@unibo.it

**Abstract.** As Europe is experiencing a second violent CoVid-19 storm, with the PCR-based testing system deteriorating due to the high volumes of people to be tested daily, there is a general reconsideration of the mathematical theories at the basis of our contact tracing and testing approaches. Drawing upon the concept of super spreader, we propose the use of (less sensitive) rapid tests to detect those secondary infections that do not need the use of PCRs, thus saving the most part of PCR tests currently used. This before the system fails.

## 1    Introduction

There is much unexplained about how CoVid-19 spreads all over the world, leaving many questions unanswered [1]. Just to cite a few: Why three neighboring regions in Italy have paid the most part of the national death toll, with almost the totality of these deaths concentrated in the first months of the initial Italian outbreak? Did the Sweden's approach succeed or fail, with its excessive death rate during the first wave, and the current low case count, as the rest of Europe experiences a second wave? Why was just one patient, in Daegu (South Korea), the cause of more than 5,000 known infections, while nothing similar had happened in Seoul, which is only 200 kilometers away? Indeed, many computational/mathematical models and theories have been used to give answer to those questions, with the final aim to predict resurgences/decays of this virus; yet none of them has been able to explain neither the scale nor the timing of these disastrous events [2]. Traditionally, epidemiological models dominate this scenery. They compute the number of people infected by a contagious disease in a closed population, considering the passage of time. The actual value of these models is out of question, nonetheless they turned out to be slow in the capability of providing early signals of uptrend for new infections and deaths. More recently, often presented as a state of the art advancement, new intelligent approaches, based on the use of machine/deep learning techniques, were proposed that aggregate data to anticipate when the virus returns or decays. Again, doubts have been raised with regard to the precision and the reliability with which also these models describe a pandemic's spread [3, 4]. Given these general failures, many researchers all over the world are wondering if all the knowledge corresponding to ten months of epidemiological data have usefully influenced the way we think about the models used to describe this pandemic diffusion [5]. With this regard, a new way to look at this is emerging. It is based on a novel interpretation of the concept of the so-called *reproductive number*. Let us refer to it, just for the purposes of the present article, using the letter *r*. Essentially, *r* is a measure with which we try to count how much contagious a virus is, *on average*. Simply put, *r* returns the *mean number* of people expected to become infected, if they are exposed to someone already hit by that virus. With the influenza (or flu), for example, *r* possesses the virtue of being key to interpret how that disease spreads. And for good reason: in that case, *calculating the average* is useful to represent the statistical distribution of that phenomenon. But, if we consider the opposite case, when the number of persons who becomes infected is distributed with an over-dispersion with respect to its mean value, then everything comes reshuffled. This looks like to be the case, here. Many recent scientific investigations, in fact, are suggesting that as few as 10/20% of infected people are responsible for as much as 80/90% of the transmission of the infection to other people. And, on the converse, the most part of infected people seem do not have any responsibility in the transmission of this disease. This article presents some reflections on what we have lost, in terms of comprehension of the phenomenon, having neglected this statistical dispersion around the *average*. Further, following this line of reasoning, we also discuss what we should change, in practical terms of contact tracing and testing, if we stop thinking that this disease spreads following a deterministic, linear and easily predictable trajectory [6]. The reminder of the paper is structured as follows. The next Section provides a basic background for introducing the concept of super spreading. Sections 3 illustrates a new audacious method for conducting the contact tracing and testing activities, while Section 4 concludes the paper.

## 2 Reproductive Number, Super Spreading and Negative Binomial

We begin by simply reminding that with the parameter *r* (or *R0* or *Rt*, respectively *basic* or *effective*, in the specializing epidemiological literature) we indicate the reproductive number, that is the number of infections which is caused, on average, in a population by an infected person. We also know that in a deterministic spread of an epidemic, this is a very important indicator to measure the evolution of the underlying disease. For instance, in commonly used infection models, when *r* is larger than 1, this means that the infection will be able to start spreading, while it can be considered under control as soon as this number *r* is brought back to values smaller than 1. Not only, but the larger this value is, the more complex it is to devise containment measures that can control the diffusion of the epidemic, up to the point when *r* is (much) larger than two, as in this case we observe an exponential growth of the number of the new infection cases.

As to CoVid-19, unfortunately, many recent studies have reported multiple cases where a single person has infected 80% or more of the people in a given place, while it has also been found that a lot of infected people may not infect a single other person. This is exactly where the statistical dispersion comes in, along with the concept of *super spreading*. Essentially, with *super spreading* we refer to a situation where some given individuals, disproportionately, infect a large number of secondary cases, while instead an average infectious individual should just infect *r* new persons. Epidemiologists have already had experiences with kind of epidemic diffusion pattern, with other recent diseases, like the first SARS, and then MERS, and even Ebola. From a mathematical viewpoint, it is obvious that the models capable to take into account this kind of behavior differ sharply from average-based approaches (based on the use of the unique *r* parameter). Hence, when super spreading situations become the main mode on the basis of which a disease spreads, and not only rare events, the relative mathematical models, called to describe how an epidemic grows, should change. In those cases, under the assumption that the number of generated secondary infection cases is over dispersed with respect to a mean value, they come to help different mathematical representations, typically based on the use of a *negative binomial distribution*, where the variance, that is the measure of the dispersion of many real cases, with respect to the mean value *r*, is computed with the following formula: *variance =* (*r* times (1 + *r/k*)) [7, 8, 9, 10]. And, here, arrived on the scene is finally a new actor: the *k* parameter. If we look at these facts in a simplistic way, we could say that when *k* is less than 1, this implies a substantial super spreading, up the point when a low value for *k* corresponds to a higher degree of super spreading. To explain better about *k*, you should think at its role in a negative binomial distribution. A traditional way to explain the role that a negative binomial distribution plays is that of comparing with a dice game, where if we set the number 1, on a given die, as a *failure*, while any other number is a *success*, and we ask how many successful rolls we will need to observe, before we have, for example, the fourth failure (*k* = 4), this is exactly the case when this kind of distribution describes the number of *non-1*s that we will observe. Translated into epidemiological terms, a negative binomial probability distribution can better capture an infectious disease transmission, where the likely number of secondary infections may vary remarkably, depending on the specific individuals that propagate the infection (are they super spreaders or not?), as well as on the setting where those individuals play its role. And then we come to the final question of this Section: How a person can become a super spreader within the context of CoVid-19? Answering this question is extremely difficult, as many factors can give a relevant contribution (including bad luck), that are still to be discovered and can influence super spreading, yet what we already know is that a necessary condition, most of the time, for a super spreading event to happen is: many people, no distancing, no mask, a prolonged contact, a poorly ventilated indoor place.

## 3    Contact Tracing, Rapid Tests and Beyond

After several months of observations of what an over-dispersed pathogen can make, we have learnt that we are not facing an infectious phenomenon which is linear and predictable. With CoVid-19, we can move very fast from one relatively safe state to another, much more complex to manage. To be more practical and useful, we should inform that it is of vital importance to develop methods that figure out when super spreading happens, using appropriate contact tracing and testing regimes. Before proposing a new method, based on an alternative to traditional PRC tests, we briefly introduce some basics about the testing methodology [10-17]. To begin: each testing method bases its efficacy on the two figures of merit: *sensitivity* and *specificity*. If tests are highly *sensitive* this means that they are good at identifying people who are infected, while if they are specific this means that they good at identifying not infected people. PCR tests have been designed to be highly accurate over both the dimensions (i.e., sensitivity >90%, specificity >80%). The problem, nonetheless, is that PCR tests are slow, expensive, and uncomfortable. Not only, but consider, for example, the situation in Italy, as of the mid of November 2020 when we have achieved as many as 30/40.000 infections per day. Counting a number of contacts of ten people to test per each new infection, we yield as many as 300/400.000 PCR tests to perform, daily. This is not sustainable. So many infection chains go lost, with an impact on the diffusion of the virus. An error here is that, based on identifying people with sanitary requirements, test manufacturers have prioritized sensitivity. Instead, by slightly relaxing those guidelines to 70, 60 or even 50% sensitivity values, we could have rapid, cheap and comfortable (e.g., saliva based) tests, that come with a specificity of 90%. Some manufacturers are already producing those tests, able to discover 60% of positives and 90% of negatives. Let us now show how these rapid tests could be profitably utilized in a situation where we are not able to sustain a load of 3/400.000 PCRs per day. Here come to help the super spreading theory: if a super spreader is not that easy to be detected, an infected person who has not super spread the virus could be easily detected, instead. Assume that if an infected people is not a super spreader this means that out of his/her 10 contacts 9 should not be infected, too. With a super spreader, we have the opposite situation: 9 infected and 1 non-infected people. Consider now a rapid test with the following characteristics: specificity: 90% and sensitivity: 60%. This corresponds to the following conditional probability values: $prob\{$**neg-test | neg**$\} = 0,9$; $prob\{$**pos-test | neg**$\} = 0,1$; $prob\{$**pos-test | pos**$\} = 0,6$; $prob\{$**neg-test | pos**$\} = 0,4$. Obviously, the conditional probabilities above hold, in general, but our goal is to decide whether a primary infection is either a super spreader or not, judging on the basis of his/her ten contacts. Hence, we calculate the probability that a person is positive/negative, given that its test has resulted as negative/positive, based on either the case when the number of contacts achieves the 90% of negatives, or 90% of positives. Call these probabilities $Pp/Tn$ (and $Pn/Tn$) and $Pp/Tp$ (and $Pn/Tp$), while with $Pn = 0,1/0,9$ and $Pp = 0,9/0,1$, we denote the *a priori* probability for an individual of being negative and positive, given he/she belongs to a chian of either a super spreader or not. The formula for calculating $Pp/Tn$ ($Pn/Tn$ can be obtained as 1 - $Pp/Tn$) and similarly $Pp/Tp$ (and $Pn/Tp$) can be obtained as follows, using both the law of total probability and the Bayes theorem:

$Pp/Tp = prob\{$**pos | pos-test**$\} = prob\{$**pos and pos-test**$\} / prob\{$**pos-test**$\} =$
$\quad = (prob\{$**test-pos | pos**$\} * prob\{$**pos**$\}) /$
$\quad\quad / ((prob\{$**pos-test | pos**$\} * prob\{$**pos**$\}) + (prob\{$**pos test | neg**$\} * prob\{$**neg**$\}))$.

The values for: $Pp/Tp$ ($Pn/Tp$) and $Pn/Tn$ ($Pp/Tn$), are reported in Table 1:

**Table 1.** *Pp/Tn*, *Pn/Tn*, *Pp/Tp*, *Pn/Tp*, for Super Spreader and Not Super Spreader

| Value | Super Spreader | Not Super Spreader |
|---|---|---|
| *Pp/Tn* | 80% | 5% |
| *Pn/Tn* | 20% | 95% |
| *Pp/Tp* | 98% | 40% |
| *Pn/Tp* | 2% | 60% |

Given the above values, the method could go as follows. For each primary infection, we take one of his/her 10 contacts and we subject him/her to a rapid test. If the test returns positive, we deliver that contact, along with all the other 9, to a PCR testing activity. This because the probability that this test has returned a true positive is very high under the hypothesis of a super spreader (98%), and not negligible, even in the case of a non-super spreader (40%). Instead, if the test is negative we could decide that any further PCR testing activity could be avoided, yet with a probability that that test has produced a false negative, which is negligible in the case of a non-super spreader (5%), but very high in the case of a super spreader (80%). In summary, we would have avoided to perform 10 PCR tests (out of ten contacts), yet with an 80% of probability we have taken the wrong decision, if we have a super spreader behind. Consider, now, to extend this rapid testing activity to all the ten contacts of a primary infection, and assume we get ten negative results, on the basis of which we decide that no PCR testing activity should be performed. In such a case, we would have lowered the probability we are taking a wrong decision (if a super spreader lies behind) from 80% to almost 10%, yet at the cost of having also lowered (from 95% to almost 60%) the percentage of true negatives, in the case of a non-super spreader. Thus increasing the number of times we have to deliver some chains of contacts to the traditional PCR testing activity. In summary, considering an incidence of a 20% of super spreaders out of 1000 new infections, with our method, we would perform the following total number of PCR tests: (200 – (200 * 10/100)) + ((800 * 40)/100) = 500, instead of 1000. This at the cost of 1000 rapid tests, plus the fact that ten percent of super spreaders go lost (i.e., 20 out of 200).

## 4    Conclusion

After an introduction on super spreading, we have proposed a new method to conduct testing activities based on rapid tests that halves the number of required PCR tests, the main drawback being that the 10% of super spreaders is lost. This should be considered, not only as a viable alternative to the current testing activity which is close to a failure given the increasing number of the new primary infections, but also a starting point from which reasoning about rapid tests.